\documentclass[
    ,final            
  ]
  {aipproc}

\layoutstyle{6x9}

\usepackage{epsfig}
\usepackage{graphicx}
\usepackage{bm}

\newcommand{\be}{\begin{equation}}

\newcommand{\ee}{\end{equation}}
\newcommand{\bea}{\begin{eqnarray}}
\newcommand{\eea}{\end{eqnarray}}

\begin{document}

\title{Indications of a Pseudogap in the\\ Nambu Jona-Lasinio model}


\author{Paolo Castorina}{
  address={Department of Physics, University of
Catania, Italy and INFN-Catania, Italy.}
}

\author{Giuseppe Nardulli}{
address={Department of Physics and TIRES Center, University of
Bari, Italy  and  INFN-Bari, Italy \\
and  PH Department, TH Unit, CERN, 1211 Geneva 23, Switzerland.
}
}

\author{Dario Zappal\`a}{
  address={INFN- Catania, Italy and Department of Physics, University of
Catania, Italy.}
}

\begin{abstract}

The survival of $\bar q q $ bound states at  temperatures higher than the 
chiral restoration temperature, $T_c$, recently observed in lattice QCD, is 
discussed in the framework of the Nambu Jona-Lasinio model.
The perturbative determination of the spectral function provides
an indication of a pseudogap phase above $T_c$.
\end{abstract}

\maketitle

\section{Introduction}

Recent lattice studies about the Quantum Chromodynamics transition at finite temperature
from the hadronic to the deconfined phase, show that, heavy, and possibly also light mesonic 
bound states survive up to twice the deconfinement temperature (for a recent review see \cite{karsch2} and references therein),
in contrast with the former  suggestion that a strong suppression  of heavy bound states, such as the
$J/\psi$, would occur just above the critical temperature. 
This unexpected feature indicates
some not yet understood mechanism at the deconfinement transition.
In addition, the  first experimental results collected at RHIC  support the presence of 
a strongly interacting phase above the critical temperature.

This picture has some analogies with the physics of high temperature superconductors where 
the coherence length is much smaller than in ordinary superconductors and 
the temperature corresponding to the onset of superconductivity turns out to be much lower than 
the temperature related to the Cooper pairs formation. Between these two  temperatures 
a pseudogap  is observed, which consists in a depletion of the single particle density of states around 
the Fermi level.  These  features could be regarded as a phenomenological manifestation of 
a crossover from the ordinary Bardeen-Cooper-Schrieffer superconducting behavior to
a Bose-Einstein Condensate behavior. 

An explanation of the high temperature superconductivity has been provided by Emery and Kivelson 
(see \cite{Emery:1995mn}) who suggested that phase fluctuations of the condensate should be responsible
for the spoiling of the long range coherence, the appearance of the pseudogap  and the
consequent  lowering of the temperature  associated to the onset of superconductivity . 
The relation between the pseudogap transition and field fluctuation has been explored 
within the framework of the Nambu Jona-Lasinio (NJL) model \cite{vari1,vari2,vari3}, and also indicated
as a possible explanation of the observed lattice bound states above the critical temperature\cite{cnz}.

Following \cite{cnz},  after a brief analysis of some results in mean field theory for the NJL model,
we briefly discuss the effects of the fluctuations and the consequent appearance of two 
characteristic temperatures:  $T_c$, corresponding to the restoration of chiral symmetry for the NJL model,
and $T^*$ associated to the decoupling of the mesons $\bar q q $ from the spectrum.
Finally we
present a calculation of the pseudogap between $T_c$ and $T^*$. 

\section{NJL model in the mean field approximation}

We consider the NJL model \cite{njl1,njl2} with an isospin doublet of massless quarks  with three colors ($N_f=2$, $N_c=3$ )
at finite chemical potential $\mu$
\be\label{uno}
{\cal L}=\bar\psi(i\partial_\nu\gamma^\nu+\mu\gamma_0)\psi
+\frac{G_0}{2N_c}\left[({\bar \psi}\psi)^2+({\bar
\psi}i\gamma_5{\tau}\psi)^2\right] \; .
\ee
As it is well known (for reviews on the NJL model see e.g. \cite{kle1},\cite{kuni1}), 
the mean field approximation provides the self-consistent equation 
(at $\mu=0$)
for the mass gap $m_q$:
\be\label{due}
m_q=4N_fN_c\frac{G_0}{2N_c}\int_\Lambda
\frac{d^3p}{(2\pi)^3}\frac {m_q}{\sqrt{p^2+m^2_q}}
\ee
where  $\Lambda$ is a 3D cutoff.
Moreover the pion decay constant $f_\pi$, again at $\mu=0$, is given by
\be\label{tre}
 f_\pi^2=-4im_q^2N_c\int_\Lambda\frac{d^4p}{(2\pi)^4}\frac{1}
{(p^2-m_q^2+i\epsilon)^2}
\ee
and one can use Eqs. (\ref{due}) and (\ref{tre}) to get $\Lambda$ by fixing $m_q$.
In particular we use $m_q=300$ MeV and $f_\pi=93.3$ MeV as input, which gives $\Lambda=675$ MeV.

The approach is then easily generalized at finite temperature and density to obtain a new gap equation for
$m_q(T,\mu)$ which, after getting rid of the coupling $G_0$ by means of Eq. (\ref{due}),  reads
\be\label{quattro}
0=\int_\Lambda\frac{d^3p}{(2\pi)^3}\left[\frac{1}{\sqrt{p^2+m^2_q}}
-\frac{\sinh\,y}{\epsilon\left(\cosh\,y+\cosh
\,x\right)}\right]
\ee
where $\epsilon=\sqrt{p^2+m^2_q(T,\mu)} $ and
$x=\mu/T$ and  $y={\epsilon}/{T} $. 

The gap equation at finite temperature and density naturally provides  a critical line signaling a phase transition. 
In fact, the  vanishing of the solution of   Eq. (\ref{quattro}): $m_q(T,\mu)= 0$ defines the critical temperature   $T^*(\mu)$ and 
its plot  in the $T-\mu$ plane is displayed in fig. 1, with the input for the  3D cutoff $\Lambda$ obtained from the values of 
$m_q$ and $f_\pi$ at $T=\mu=0$ given above.
Our next step is to  check to what extent the fluctuations can modify the results of the mean field analysis.

\begin{figure}
\includegraphics[height=.25\textheight]{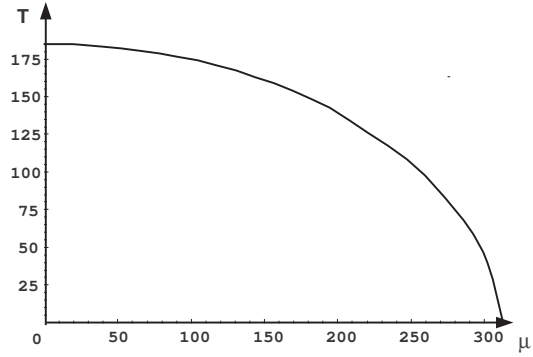}
\caption{ Plot of $T^*(\mu)$  in MeV, as obtained from Eq. (\ref{quattro})  for $\Lambda=675$ MeV. }
\end{figure}

\section{Fluctuation effects}

In  order to study the modifications to the mean field analysis it is necessary to introduce a method
that can properly account for the fluctuation effects. Such approach has been developed 
\cite{vari2,vari3} and its essential features will be briefly illustrated in the following.

The four fermion interaction in the model in Eq. (\ref{uno}) can be cast in the form of Yukawa coupling,
by introducing the auxiliary scalar and pseudoscalar fields $\sigma$ and $\vec \pi$: 
\be\label{cinque}
{\cal L}=
\bar\psi(i\partial_\nu\gamma^\nu+\mu\gamma_0-g_0(\sigma+i\gamma_5{\vec
\tau}\cdot{\vec\pi}))\psi-\frac{g_0^2N_c}{2G_0}\left[\sigma^2+{\vec
\pi}^2\right]
\ee
with the quark-meson coupling constant given by $g_0=m_q/ f_\pi$, which is the analogous of the 
Goldberger-Treiman relation. By functional integration of the fermionic degrees of freedom  and derivative  
expansion, Eq. (\ref{cinque}) becomes equivalent to the $\sigma$ model with lagrangian 
\be\label{sei}
{\cal L}_\sigma=\frac\beta 2\left(
(\partial\sigma)^2+ (\partial{\vec
\pi})^2-\frac{\kappa^2}4\left(\sigma^2+{\vec\pi}^2-f_\pi^2\right)^2\right)
\ee
where the stiffness parameter, $\beta$, which is relevant for our purposes, is given by \cite{vari2,eguchi}
\be\label{sette}
\beta=4g_0^2N_cN_f \left (
\int_\Lambda \frac{d^4p_E}{(2\pi)^4}\frac 1{(p_E^2+m^2_q)^2} -
\frac 1 2 \int_\Lambda \frac{d^4p_E}{(2\pi)^4}\frac
{p_E^2}{(p_E^2+m^2_q)^3} \right )
\ee
and $p_E$ is the Euclidean four-momentum.

The key point is that an equivalent description of the problem  should be obtained 
by resorting to an effective theory, defined by a non-linear $\sigma$ model 
for the fields $\sigma$ and $\vec \pi$, satisfying the constraint
$\sigma^2+{\vec\pi}^2\,=\,f_\pi^2$ \cite{vari2,vari3}. This constraint can be introduced into the functional generator $Z$
of the non-linear $\sigma$ model, by means of a functional Lagrange multiplier $\lambda(x)$ 
\be\label{otto}
Z=\int [d\sigma][ d{\vec\pi}][d\lambda]\exp\left \{i
\frac{\beta}2\int
d^4x\left((\partial\sigma)^2+(\partial{\vec\pi})^2-\lambda
\left(\sigma^2+{\vec\pi}^2-f_\pi^2\right) \right) \right \}
\ee
The parameter $\beta$ in Eq. (\ref{otto}) indicates the stiffness of the non-linear $\sigma$ model.

The functional integration of the $\vec \pi$ fields in Eq. (\ref{otto})  can be performed and, 
afterward, one can use the saddle point approximation and search for $x$-independent solutions for 
$\sigma$ and $\lambda$. As it is evident from Eq. (\ref{otto}), the constant $\lambda$
plays the role of a square mass both for $\sigma$ and $\vec \pi$ fields.
Since we are looking for phase transitions at finite $T$ and $\mu$, we consider 
Matsubara frequencies $\tilde\omega_n=2\pi nT$ and non-vanishing baryonic 
chemical potential and the two saddle point conditions read
\be\label{nove}
0=\lambda\sigma \; ;   \hskip 30 pt 0=\beta(f_\pi^2-\sigma^2)-(N_f^2-1)
\,\sum_{n=-\infty}^{\infty} \int_{\Lambda_\pi}   \frac{d^3  k}{(2\pi)^3} \frac
T{(\tilde\omega_n-i\mu)^2+k^2+\lambda}
\ee
where $\Lambda_\pi$ is a suitable cutoff for the non-linear $\sigma$ model.

The first condition $0=\lambda\sigma $ indicates that at least one of the two variables must vanish
and the other variable is  implicitly determined 
from the second condition in Eq. (\ref{nove}),
in terms of $T$, $\mu$ and of the stiffness
$\beta$.
The equivalence of the non-linear $\sigma$ model with the lagrangian in Eq. (\ref{sei}) implies that 
the stiffness evaluated in the two cases must coincide. Therefore one can insert the 
expression of $\beta$ of Eq. (\ref{sette}), properly modified to account for the finite $T$ and $\mu$ effects, i.e.
\be\label{dieci}
\beta=2g_0^2N_fN_c
T\sum_{n=-\infty}^{+\infty}\int_\Lambda\frac{d^3p}{(2\pi)^3}\left[\frac
1{[\epsilon^2+(\omega_n-i\mu)^2]^2}+\frac{m^2_q(T,\mu)}{[\epsilon^2+
(\omega_n-i\mu)^2]^3}\right]
\ee
into Eq. (\ref{nove}) and determine the corresponding solution for $\sigma$ or $\lambda$.

As an example, in fig. 2 we plot the solutions of Eq. (\ref{nove}):
$\sigma$ and the mass of the pionic mode $m_\pi=\sqrt{\lambda}$
versus  the temperature, at $\mu=0$ and for two different values of 
the number of colors: $N_c=3$ (solid lines) and $N_c=10$ (dashed lines).
The values of the two cutoffs entering Eqs. (\ref{nove}) and (\ref{dieci})
are $\Lambda_\pi=200$ MeV and $\Lambda=675$ MeV. 
In fact  $\Lambda_\pi$ and  $\Lambda$ are related to different degrees 
of freedom and do not need to have the same value and we take a value of $\Lambda_\pi$
close to $f_\pi$, which fixes the scale of the non-linear $\sigma$ model. 
(The choice $\Lambda_\pi\sim\Lambda$ 
made in \cite{vari1} has been criticized in \cite{vari2}. For other references on this point see \cite{cnz}.)

\begin{figure}
  \includegraphics[height=.3\textheight]{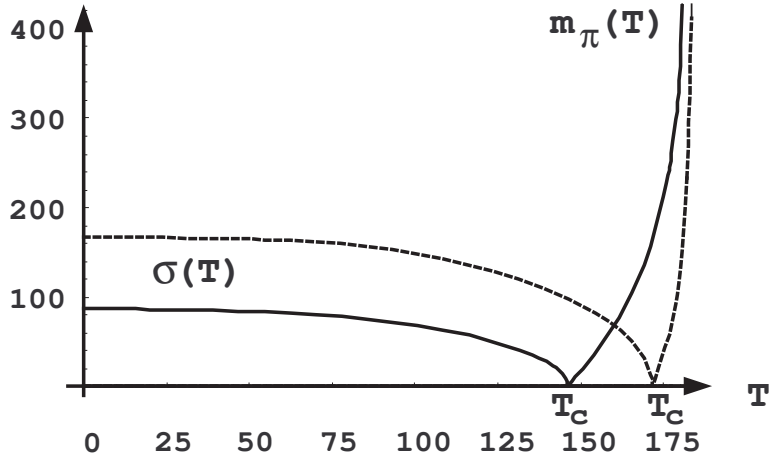}
\caption{$\sigma$ and $m_\pi$ vs.  $T$ at $\mu=0$ (in MeV) as obtained from Eq. (\ref{nove}). See
text for the specific values of the other parameters.}
\end{figure}

We observe in fig. 2 that at low temperatures the solution 
of Eq. (\ref{nove}) corresponding to $\lambda=0$ (not plotted) and 
non-vanishing $\sigma$  is realized. Then $\sigma$ decreases with $T$ 
and eventually vanishes. By further increasing the temperature,
the solution  $\sigma=0$ (not plotted) and $m_\pi=\sqrt{\lambda}\neq 0$ 
is realized.

The two regions meet at $T=T_c$ where $\sigma=\lambda=0$.
This corresponds in Eq. (\ref{nove}) to the critical stiffness
\be\label{undici}
\beta_c=\frac{N_f^2-1}{f_\pi^2(T_c,\mu)}
\,\sum_{n=-\infty}^{\infty}\int_{\Lambda_\pi} \frac{d^3
k}{(2\pi)^3}\frac {T_c}{(\tilde\omega_n-i\mu)^2+k^2}
\ee
and $T_c$ is determined by   the comparison of Eqs. (\ref{dieci})  and (\ref{undici})
In the particular case considered in fig. 2 we get  $T_c\approx 146$ MeV for 
$N_c=3$ and $T_c\approx 172$ MeV for $N_c=10$.

One should notice that in the non-linear $\sigma$ model the region below $T_c$ 
shows a non-vanishing expectation value of the field $\sigma$ which implies  
a dynamical breaking of the chiral symmetry. Above $T_c$ this expectation value 
vanishes and at the same time a finite mass for the pions is generated.
The absence from the spectrum of massless Goldstone  bosons is a signal
of chiral symmetry restoration.  Therefore, within this framework, $T_c$ is
the critical temperature related to the chiral symmetry.

In fig. 2, when the temperature is increased,  $m_\pi$ grows and, 
at some temperature, $T_{pair}$, eventually diverges and 
the $\vec \pi$ mesons decouple and disappear  from the spectrum.  Reasonably, 
$T_{pair}$ could be interpreted as the  temperature  of dissociation
of the $\bar q q$ pair. A remarkable feature which appears in fig.2 is that 
$T_{pair}$  does not depend on $N_c$  and, in practice, it numerically  
coincides with the critical temperature $T^*$ determined by the mean field analysis:
 $T_{pair} \sim T^* \sim 185$ MeV.  

Therefore, the comparison of the NJL model with the non-linear 
$\sigma$ model indicates that   fluctuations indeed modify the mean field conclusions 
and the new critical temperature $T_c$, associated with the chiral symmetry restoration, 
is generated. $T_c$ turns out to be smaller than $T^*$ so that we find a temperature
interval  $T_c<T<T^*$ where the gap equation still provides $m_q\neq 0$
but the chiral symmetry is restored as an  effect of the fluctuations. Only  above $T^*\sim T_{pair}$ 
the mesons disappear and $m_q$ vanishes. Consistently with this picture we note  
that $T_c$ depends on $N_c$ and $T_c \to  T^*$  when  $N_c\to \infty$, i.e. when 
the fluctuation effects become   suppressed and the mean field regime holds.

A comment about $m_q$ is in order. In fact the common belief is that a non-vanishing $m_q$ 
is directly related to the breakdown of chiral symmetry. This is true only if the  gap $m_q$ 
appears as a pole in the two point  function of the fermion $\psi$. In our  specific case 
$m_q$ corresponds to the pole mass of $\psi$ only in the mean field approximation.
This property has been first  observed  in 2D models in \cite{witten},
and, for a  detailed treatment of the problem, we refer to  \cite{nikolov,loktev}.
With a suitable change of variables, corresponding to the introduction of 
polar 'coordinates':  $\sigma + i \gamma^5 \pi =\rho~{\rm exp }(i \gamma^5 \theta)$ and of 
the fermion $\chi= {\rm exp }(i \gamma^5 \theta/2)$, where $\rho$ and $\chi$ are chirally neutral,
it is possible to show that  $m_q$, as obtained from the gap equations (\ref{due})
and (\ref{quattro}) is a pole in the $\chi$ two point function (of course in our specific case the definition of the new variables 
should be modified to account for the flavor degrees of freedom).
Due to the chiral neutrality of $\chi$,  this excitation does not break chiral symmetry. 
Conversely, it does not  appear in the $\psi $ propagator  as a pole but rather, 
due to the  $\theta$ field fluctuations, as a branch cut. In this sense, a finite $m_q$ is still 
compatible with chiral symmetry.

The presence of a region between two characteristic temperatures 
where the fluctuations cancel the long range order associated to the quark mass,
restoring the symmetry,  shows a clear resemblance with the pseudogap 
phase of high temperature superconductors \cite{vari1,vari2,loktev} and this motivates 
the following analysis of the spectral function. 

\section{Evaluation of the pseudogap}

After having determined the two characteristic temperatures $T_c$ and $T^*$, we conclude by showing 
that a simple  approximation in the calculation of the spectral function provides clear indications of a pseudogap \cite{cnz}.
The spectral function  of the fermionic quasiparticles, $N(\omega)$,  is obtained from the imaginary part of $G^R({\vec k}, \omega)$:
\be\label{dodici}
N(\omega)=- \frac{1}{\pi}\int\frac{d{\vec k}}{(2\pi)^3} {\rm Tr}_{c,f,d}~~
\gamma^0  ~{\rm Im} ~G^R({\vec k}, \omega)
\ee
where $ {\rm Tr}_{c,f,d}$ is the trace over Dirac, color and flavor degrees of freedom.
$G^R({\vec k}, \omega)$ is the analytical continuation of the imaginary time retarded Green function:
$G^R({\vec k}, \omega)=\left[G^{-1}({\vec k},\omega+ i \epsilon)-\Sigma_R({\vec k},\omega) \right]^{-1}$
where $G^{-1}({\vec k},\omega+ i \epsilon)$ corresponds to the free term  contribution and  $\Sigma_R({\vec k},\omega)$
includes higher order corrections.
The starting point to calculate the retarded Green  function is the Lagrangian 
\be\label{quindici} 
{\cal L}_{eff}=
\bar\psi[i\partial_\nu\gamma^\nu+\mu\gamma_0-g_0(\sigma+i\gamma_5{\vec
\tau}\cdot{\vec\pi})]\psi-\frac{g_0^2N_c}{2G_0}\left[\sigma^2+{\vec
\pi}^2\right]
\ee
where the $\sigma$  field is replaced by expanding the constraint :
$\sigma= \left ( f_\pi^2 -{\vec\pi}^2\right)^{1/2}  \sim f_\pi -   {\vec\pi}^2/(2 f_\pi) $.
The correction  $\Sigma_R({\vec k},\omega)$ to the free fermion propagator is computed   
to the one loop level from Eq. (\ref{quindici}), \cite{eguchi}, and its imaginary part generates
the correction, $N_{pert}$,  to the free fermion contribution  in  the spectral function 
\be\label{sedici}
N(\omega)=\frac{N_cN_f} {\pi^2}(\omega+\mu)\sqrt{(\omega+\mu)^2 -m_q^2}+
N_{pert} \; .
\ee
In fact we have included the  mass $m_q$, which is dynamically generated, in  the free fermion contribution 
and we have also included in the definition of $g_0$ the form factor 
$g_0^2({\vec  q}^2)=g_0^2\,{m_P^4}/ \left ({m^2_P+(2\pi nT)^2+{q}^{2}}\right)^2$
with $m_P=100$ MeV, which acts as a regulator that corrects the large momentum behavior of the loop.

We normalize $N(\omega)$ to the free massless case : $ R(\omega) =N(\omega)/ ( \frac{N_cN_f}{\pi^2}(\omega+\mu)^2) $
and plot  $R(\omega)$  for $(T-T_c)/T_c=0.02,0.15,0.20$  ($\mu=10$ MeV) in fig. 3. The solid straight line corresponds to $T=T^*$.
In fig. 3 we observe the pseudogap for values of $\omega$ just above the value $\omega =m_q -\mu$, where the free
part of $N(\omega)$ vanishes, whereas it disappears for large values of $\omega$. The effect of the pseudogap is
larger for temperatures just greater than $T_c$ and vanishes at $T=T^*$.
Therefore, even in this simple approximation, it is possible to realize the presence of the depletion of states  in the spectral function
and this calculation supports the picture of a pseudogap appearing between the two characteristic  temperatures. 
Since the NJL model effectively contains many ingredients of QCD, this analysis is a  starting point 
to check the existence of a pseudogap in QCD and its possible role in the explanation of the 
mesonic bound states observed in lattice calculations.

\begin{figure}
  \includegraphics[height=.35\textheight]{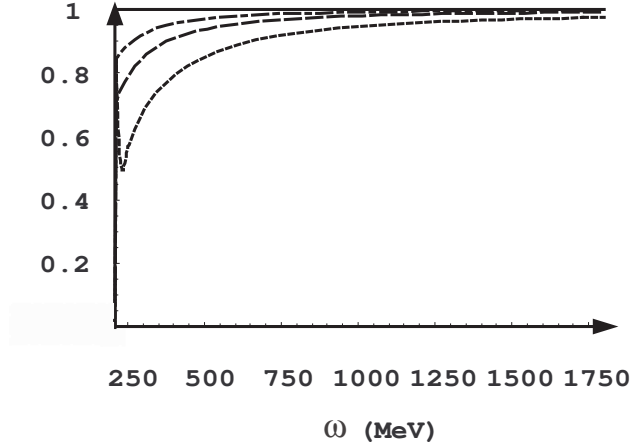}
  \caption{The ratio $N(\omega)/N_{free}(\omega)$ as a function of the energy $\omega$, at $\mu=10$ MeV and for various temperatures 
between $T_c$ and $T^*$. See text.}
\end{figure}

\bibliographystyle{aipproc}   
\bibliography{zapp}

\begin{thebibliography}{14}
\expandafter\ifx\csname natexlab\endcsname\relax\def\natexlab#1{#1}\fi
\providecommand{\enquote}[1]{``#1''}
\expandafter\ifx\csname url\endcsname\relax
  \def\url#1{\texttt{#1}}\fi
\expandafter\ifx\csname urlprefix\endcsname\relax\def\urlprefix{URL }\fi
\providecommand{\eprint}[2][]{\url{#2}}

\bibitem[Karsch(2005)]{karsch2}
F.~Karsch  (2005), \eprint{hep-lat/0502014}.

\bibitem[Emery and Kivelson(1995)]{Emery:1995mn}
V.~J. Emery, and S.~A. Kivelson, \emph{Nature} \textbf{374}, 434 (1995).

\bibitem[Hatsuda and Kunihiro(1985)]{vari1}
T.~Hatsuda, and T.~Kunihiro, \emph{Phys. Rev. Lett.} \textbf{55}, 158--161
  (1985).

\bibitem[Kleinert and van~den Bossche(2000)]{vari2}
H.~Kleinert, and B.~van~den Bossche, \emph{Phys. Lett.} \textbf{B474}, 336--346
  (2000), \eprint{hep-ph/9907274}.

\bibitem[Babaev(2000)]{vari3}
E.~Babaev, \emph{Phys. Rev.} \textbf{D62}, 074020 (2000),
  \eprint{hep-ph/0006087}.

\bibitem[Castorina et~al.(2005)]{cnz}
P.~Castorina, G.~Nardulli, and D.~Zappala  (2005), \eprint{hep-ph/0505089}.

\bibitem[Nambu and Jona-Lasinio(1961{\natexlab{a}})]{njl1}
Y.~Nambu, and G.~Jona-Lasinio, \emph{Phys. Rev.} \textbf{122}, 345--358
  (1961{\natexlab{a}}).

\bibitem[Nambu and Jona-Lasinio(1961{\natexlab{b}})]{njl2}
Y.~Nambu, and G.~Jona-Lasinio, \emph{Phys. Rev.} \textbf{124}, 246--254
  (1961{\natexlab{b}}).

\bibitem[Klevansky(1992)]{kle1}
S.~P. Klevansky, \emph{Rev. Mod. Phys.} \textbf{64}, 649--708 (1992).

\bibitem[Hatsuda and Kunihiro(1994)]{kuni1}
T.~Hatsuda, and T.~Kunihiro, \emph{Phys. Rept.} \textbf{247}, 221--367 (1994),
  \eprint{hep-ph/9401310}.

\bibitem[Eguchi(1976)]{eguchi}
T.~Eguchi, \emph{Phys. Rev.} \textbf{D14}, 2755 (1976).

\bibitem[Witten(1978)]{witten}
E.~Witten, \emph{Nucl. Phys.} \textbf{B145}, 110 (1978).

\bibitem[Nikolov et~al.(1996)]{nikolov}
E.~N. Nikolov, W.~Broniowski, C.~V. Christov, G.~Ripka, and K.~Goeke,
  \emph{Nucl. Phys.} \textbf{A608}, 411--436 (1996), \eprint{hep-ph/9602274}.

\bibitem[Gusynin et~al.(1999)]{loktev}
V.~Gusynin, V.~Loktev, and S.~Sharapov, \emph{JETP} \textbf{88}, 685 (1999).

\end{thebibliography}

\end{document}